\pdfoutput=1
\documentclass[superscriptaddress,citeautoscript,aps,prl,preprint]{revtex4-1}
\usepackage{graphicx}
\usepackage{enumitem}
\usepackage{gensymb}
\usepackage{amsmath}
\usepackage{amssymb}
\usepackage{dcolumn}
\usepackage{color}
\begin{document}

\title{2D ice from first principles: structures and phase transitions}
\author{Ji Chen}
\affiliation{London Centre for Nanotechnology, 17-19 Gordon Street, London WC1H 0AH, U.K.}
\affiliation{Thomas Young Centre, University College London, 20 Gordon Street, London, WC1H 0AJ, U.K.}
\author{Georg Schusteritsch}
\affiliation{Thomas Young Centre, University College London, 20 Gordon Street, London, WC1H 0AJ, U.K.}
\affiliation{Department of Physics and Astronomy, University College London, Gower Street, London WC1E 6BT, U.K.}
\affiliation{Department of Materials Science and Metallurgy, University of Cambridge, 27 Charles Babbage Road, Cambridge CB3 0FS, U.K.}
\author{Chris J. Pickard}
\affiliation{Thomas Young Centre, University College London, 20 Gordon Street, London, WC1H 0AJ, U.K.}
\affiliation{Department of Physics and Astronomy, University College London, Gower Street, London WC1E 6BT, U.K.}
\affiliation{Department of Materials Science and Metallurgy, University of Cambridge, 27 Charles Babbage Road, Cambridge CB3 0FS, U.K.}
\author{Christoph G. Salzmann}
\affiliation{Department of Chemistry, University College London, 20 Gordon Street, London, WC1H 0AJ, U.K.}
\author{Angelos Michaelides}
\email{angelos.michaelides@ucl.ac.uk}
\affiliation{London Centre for Nanotechnology, 17-19 Gordon Street, London WC1H 0AH, U.K.}
\affiliation{Thomas Young Centre, University College London, 20 Gordon Street, London, WC1H 0AJ, U.K.}
\affiliation{Department of Chemistry, University College London, 20 Gordon Street, London, WC1H 0AJ, U.K.}

\begin{abstract}
Despite relevance to disparate areas such as cloud microphysics and tribology, major gaps
in the understanding of the structures and phase transitions of low-dimensional water ice remain. 
Here we report a first principles study of confined 2D ice as a function of
pressure.
We find that at ambient pressure 
hexagonal and pentagonal monolayer structures are the two lowest enthalpy phases identified.
Upon mild compression
the pentagonal structure becomes the most stable and persists
up to \textit{ca.} 2 GPa at which point square and rhombic phases are 
stable.
The square phase agrees with recent experimental observations of square ice confined within
graphene sheets.
We also find a double layer AA stacked square ice phase, which
clarifies the difference between experimental observations and earlier force field simulations.
This work provides a fresh perspective on 2D confined ice, highlighting the
sensitivity of the structures observed to both the confining pressure and width.

\end{abstract}

\maketitle

Confined and interfacial water-ice is ubiquitous in nature, playing an important role in 
a wide range of areas such as rock fracture, friction, and nanofluidics \cite{chandler_interfaces_2005, xu_graphene_2010, carrasco_molecular_2012, gao_elastic_2015}.
As a result of a delicate balance of forces (hydrogen bonding, van der Waals, and interaction with the 
confining material or substrate) confined and interfacial water forms a rich 
variety of structures \cite{yang_ice_2004, carrasco_one-dimensional_2009, forster_c2_2011, nie_pentagons_2010, chen_unconventional_2014, algara-siller_square_2015}. 
Almost every specific system examined has revealed a different structure such as a 2D
overlayer built from heptagons and pentagons on a platinum surface or the 
square ice observed within layers of graphene \cite{nie_pentagons_2010, algara-siller_square_2015}.
This shows that in contrast to bulk ice the phase behavior of 2D ice is much less well understood.

From an experimental perspective a full
exploration of the phase diagram of 2D ice has not been
achieved yet.
However
recent experiments revealed the exciting possibility of exploring
2D ice structures at specific conditions by trapping
water within layered materials \cite{xu_graphene_2010, li_two-dimensional_2015, algara-siller_square_2015}.
For example by confining water between layers of graphene it is
possible to create so called nanocapillaries in which
water experiences a pressure in the GPa regime due to the van der Waals forces 
pulling the sheets together \cite{algara-siller_square_2015}.
Using transmission electron microscopy (TEM) 
square ice structures from a single up to a few layers were observed 
in such graphene nanocapillaries \cite{algara-siller_square_2015},
and in the multilayer regime square ice layers
located directly on top of one another in an AA stacking arrangement.
Force field simulations performed as part of the same study of confined water in graphene layers
reproduced the
square monolayer ice but failed to explain the AA stacking \cite{algara-siller_square_2015}.
Indeed prior to this recent study there was already a long tradition of computer simulation studies
of nanoconfined water, mostly involving classical force field approaches \cite{koga_first-order_2000, zangi_monolayer_2003, koga_phase_2005, giovambattista_phase_2009, han_phase_2010, johnston_liquid_2010, kastelowitz_anomalously_2010, bai_guest-free_2010, bai_polymorphism_2012, ferguson_computational_2012, giovambattista_computational_2012, zhao_ferroelectric_2014, zhao_highly_2014, lu_investigating_2014, corsetti_new_2015, corsetti_enhanced_2015}.
Such work has been incredibly valuable and 
provided considerable insight into the structures and phase transitions of
monolayer and multilayer ice.
The 2D ice structures predicted include hexagonal,
pentagonal, quasicrystalline, hexatic, and orthogonal phases
as well as various amorphous structures.
Although the relative stability of the structures depends on the particular
force field used,
for monolayer ice an orthogonal phase has been widely predicted to be stable across a
broad pressure regime \cite{zangi_monolayer_2003, koga_phase_2005, bai_guest-free_2010, zhao_ferroelectric_2014}.
Given the sensitivity of force field studies of confined ice to the potential
used, the aspiration to understand the observation of monolayer square ice, and the unexplained AA stacking order of
square ice layers, a systematic study with an electronic structure method such as
density functional theory (DFT) is highly desirable.
DFT studies of ice do not come without their own sensitivity to
the exchange-correlation functional used, however, they have
proved to be very useful in predicting and understanding structures of
adsorbed water and bulk ice. \cite{yang_ice_2004, tribello_blind_2006, carrasco_one-dimensional_2009, forster_c2_2011, nie_pentagons_2010, chen_unconventional_2014}.


Here we report a systematic study of 2D phases of water-ice from first principles based on an unbiased exploration of
monolayer ice structures.
The much greater computational cost of DFT compared to force fields means
that we cannot map the entire phase diagram of monolayer confined water.
Instead we focus on the phase transitions as a function of lateral pressure and confinement width at
zero Kelvin.
The stable structures identified at different pressures include an hexagonal structure, a 
pentagonal Cairo tiling (CT) structure, a flat square structure, and a buckled rhombic structure.
The observation of a flat square structure is consistent with the recent experimental observation. 
However, the sequence of low energy phases identified 
differs significantly from that predicted in force field based studies \cite{zangi_monolayer_2003, koga_phase_2005, bai_guest-free_2010, ferguson_computational_2012, zhao_ferroelectric_2014} 
and a recent DFT report \cite{corsetti_new_2015}.
Interestingly the sequence of structures observed depends sensitively on the confinement width used,
suggesting that it should be possible to tune the monolayer ice structures produced in experiments
by \textit{e.g.} varying the confining material.
We also propose a double layer square ice phase with interlayer hydrogen bonds, which further
explains the AA stacking order of multi-layer ice recently observed \cite{algara-siller_square_2015}.

\begin{figure}[h!]
\begin{center}
\includegraphics[width=8cm]{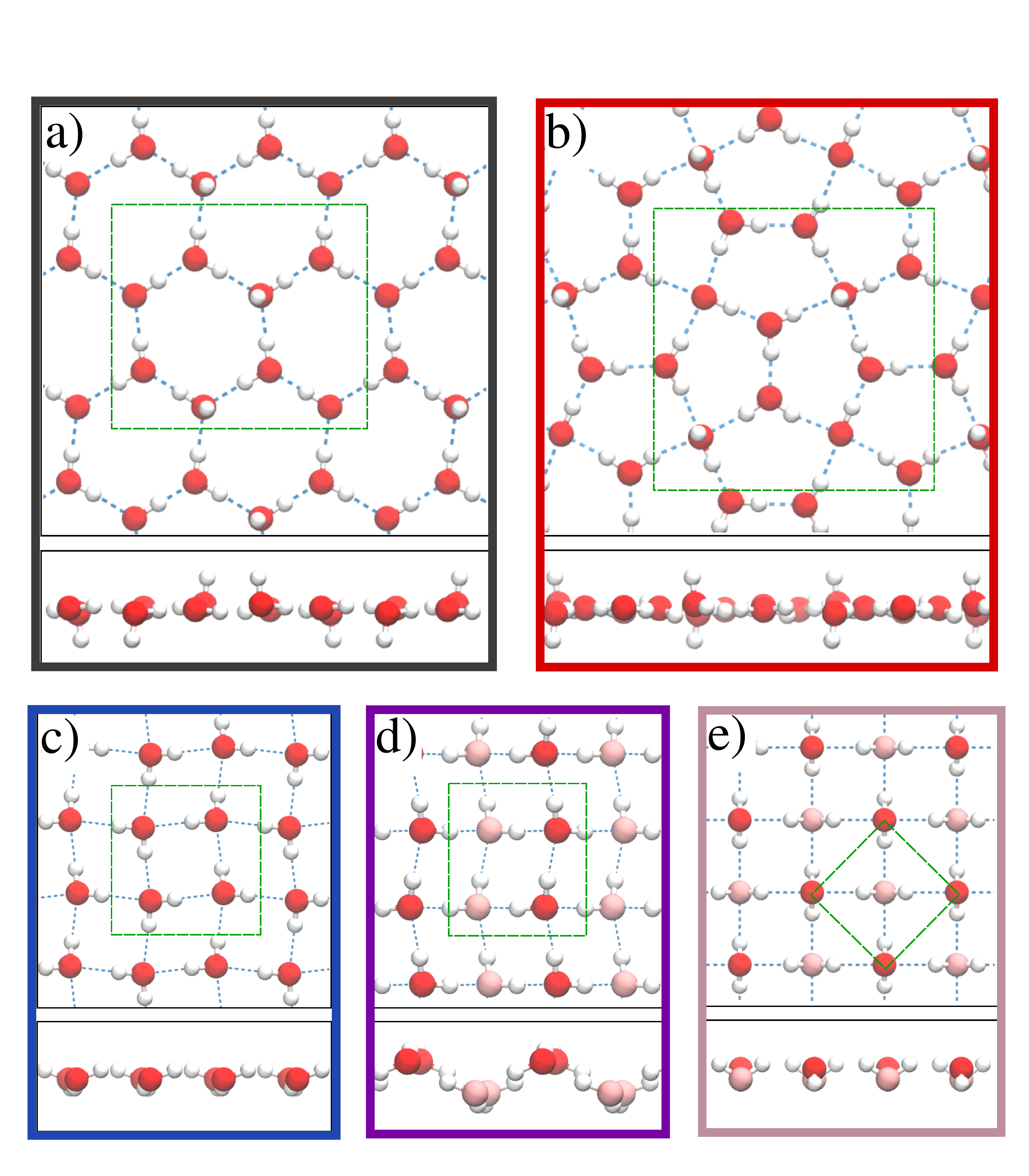}
\end{center}
\caption{Monolayer ice structures.
The top and side views of the hexagonal (a), the Cairo tiling (CT) (b),
the flat square (f-SQ) (c), the rhombic (b-RH) (d) and the buckled square (b-SQ) (e) structures.
Red and pink spheres represent oxygen atoms at different heights and white
spheres are hydrogen atoms.
The green boxes show the primitive unit cells. 
}
\label{figure1}
\end{figure}

In order to explore ice structures in an unbiased manner, we used the
 \textit{ab initio} random structure search (AIRSS) technique \cite{pickard_high-pressure_2006, pickard_ab_2011}, an approach 
which has
previously predicted new ice, 2D, and interfacial structures \cite{pickard_decomposition_2013, schusteritsch_predicting_2014}.
Structures from ambient up to a lateral pressure of
10 GPa were considered.
The 2D confinement was introduced \textit{via} a Morse potential fit to
quantum Monte Carlo results for the binding of a water monomer to 
graphene \cite{ma_adsorption_2011}.
By tuning the confinement width, we are not only able to study the general properties of
water under flat and smooth confinement but also to compare with the recent experiments 
of ice in graphene confinement \cite{algara-siller_square_2015}.
Our electronic structure calculations were carried out using the Vienna \textit{Ab-initio}
Simulation Package (VASP) \cite{kresse_efficient_1996} with the 
DFT+vdW approach \cite{tkatchenko_accurate_2009} in conjunction with the Perdew-Burke-Ernzerhof exchange-correlation functional \cite{perdew_generalized_1996}.
Tests with other exchange-correlation functionals 
show that whilst the transition pressures between the various phases depend to some 
extent on the choice of exchange-correlation functional, the overall conclusions 
do not change.
See the supporting information (SI) for these results as well as further computational details \cite{si}.

From a preliminary set of calculations we established that the optimal separation
between graphene sheets in which a monolayer of water is sandwiched is somewhere between
6.0 and 6.5 \AA~(Fig. S2) \cite{si}.
With this in mind we first report results for water within smooth confining potentials that are 
either 6.0 or 6.5 \AA~wide. 
With such confinement two phases have been identified at ambient pressure, which
have exceedingly similar enthalpies. 
These are an hexagonal monolayer structure, resembling
an hexagonal bilayer,
and a Cairo tiled structure built exclusively from water pentagons. 
With the particular exchange-correlation functional used,
the less dense hexagonal structure is marginally more stable than the pentagonal structure 
by 5 meV/$\text{H}_2\text{O}$. This difference drops to just 
2 meV/$\text{H}_2\text{O}$ when harmonic zero point energy effects are
taken into account (Table S2, Fig. S9).
Tests with other exchange-correlation functionals generally concur that
the energy difference between the two phases is tiny,
with a vanishingly small preference for the hexagonal structure (Table S2) \cite{si}.
Anharmonic zero point energy differences between the two phases 
and finite temperature effects could also easily exceed the energy
difference \cite{engel_anharmonic_2015}, suggesting that both the hexagonal
and the CT phases could be observed at ambient pressures.

\begin{figure}[h!]
\begin{center}
\includegraphics[width=8cm]{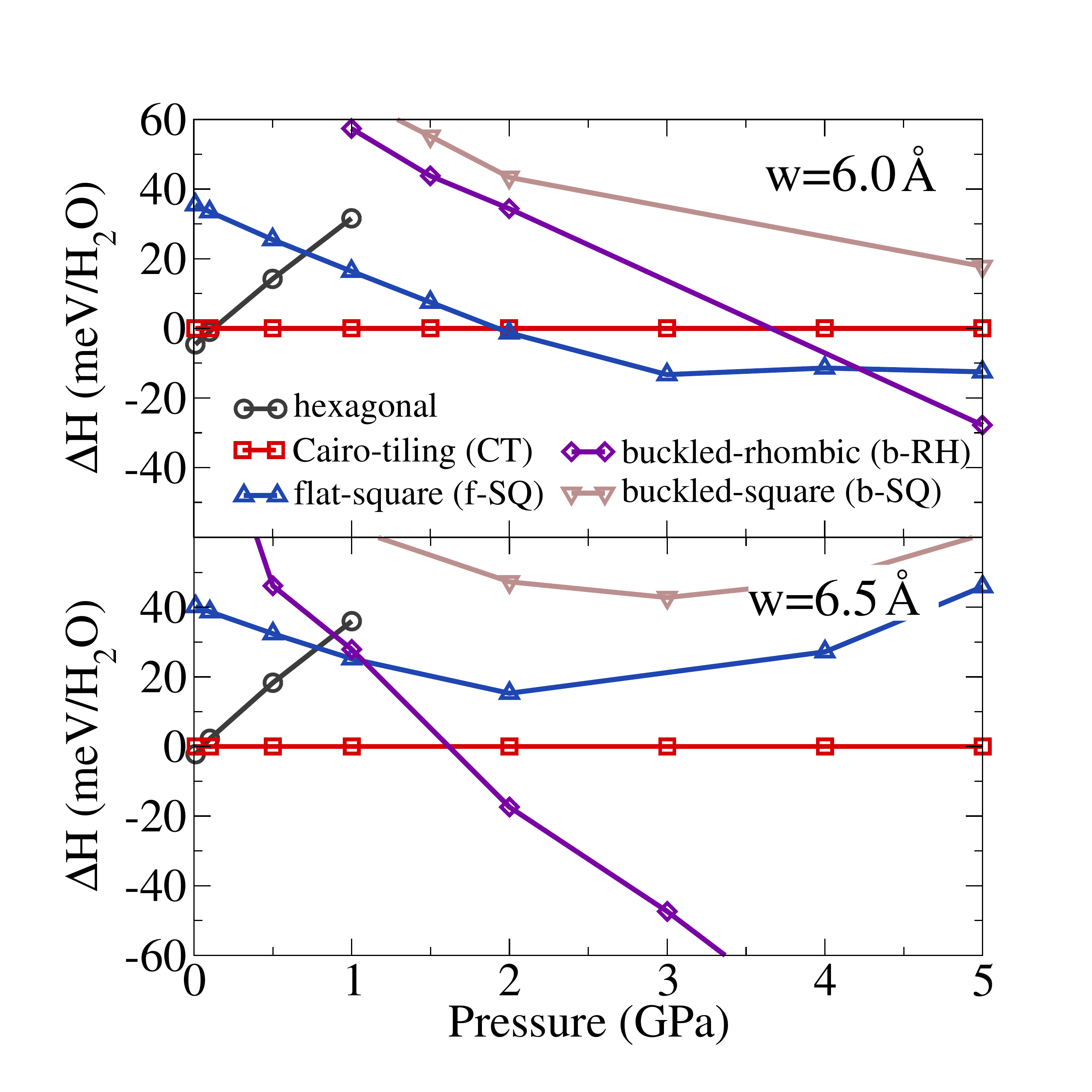}
\end{center}
\caption{Enthalpies of the water monolayer phases as a function of lateral pressure
under 6.0 \AA~(a) and 6.5 \AA~(b) confinement.
$\Delta\text{H}$ is the relative enthalpy with respect to the CT phase.
Enthalpy $\text{H} = \text{E}_\text{water} + \text{E}_\text{confinement} + \text{P} \times \text{A} \times \text{w} $,
where $\text{E}_\text{water}$ is the total energy per water molecule, $\text{E}_\text{confinement}$
is the energy (per water molecule) in the confinement potential, $\text{P}$ is the lateral pressure,
$\text{A}$ is the lateral area per water molecule, $\text{w}$ is the width of confinement.
}
\label{figure2}
\end{figure}

The hexagonal structure has \textit{p6mm} wall-paper group symmetry if only oxygen atoms
are considered and is 
built exclusively of six-membered rings (Fig. \ref{figure1}a).
Water molecules are
three-fold coordinated, with half of them having one OH bond
directed out of the monolayer (a so-called ``dangling OH").
The average O-O separation in this monolayer hexagonal phase is about 2.72 \AA~which is similar to 
the bulk O-O separation in ice I \cite{Rottger:sh0050}.
However the confined hexagonal structure identified here is quite flat with the vertical 
separation between oxygen atoms $<$ 0.3 \AA, much 
smaller than the 0.9~\AA~buckling within an hexagonal layer in bulk ice I \cite{petrenko_physics_2002}.
Given that bulk ice I is built from hexagonal layers and
double layer hexagonal structures have been observed frequently 
in force field simulations \cite{han_phase_2010, kastelowitz_anomalously_2010, johnston_liquid_2010},
it is not surprising that a low enthalpy hexagonal structure should be identified.
However monolayer
hexagonal ice has yet to be observed experimentally and 
in force field simulations it has only been found when an hexagonally patterned substrate has been used 
as a template \cite{ferguson_computational_2012}.

The pentagonal structure identified has a wall-paper group symmetry of \textit{p4gm} (Fig. \ref{figure1}b).
The unit cell has twelve water molecules, four with dangling OH bonds.
One third of the water molecules are four-fold coordinated and the rest are three fold coordinated.
The higher average coordination and smaller ring size of the CT phase renders the 
density of this phase higher than that of the hexagonal phase (Fig. S4, S5) \cite{si}.
Therefore, upon increasing the pressure the CT phase clearly 
becomes more stable than the hexagonal phase.
After searching for the lowest enthalpy structures at finite 
pressure we find that the CT structure is clearly the most stable 
in a broad range of pressures all the way up to \textit{ca.} 2 GPa (Fig. \ref{figure2}). 
The stability regime of the monolayer CT phase and the small energy difference 
at ambient pressure suggest that five-membered 
rings are more important in 2D ice than in 3D ice where they only appear
in the 0.2-0.7 GPa range (in ices III, V, IX, XIII)  \cite{salzmann_polymorphism_2011}.
There is some precedent for the pentagonal based structure being proposed here.
First and foremost, it is 
a monolayer version of the double 
layer confined ice structure 
identified in Ref. \onlinecite{johnston_liquid_2010} on
the basis of simulations with a coarse grained model of water.
In addition
a 1D pentagonal ice structure has been observed with scanning tunnelling microscopy on a metal surface,
although the structure of the 1D pentagonal chain is very different from
2D pentagonal ice \cite{carrasco_one-dimensional_2009}.
The only purely 2D structure similar to our prediction, that we are aware of,
is a recently proposed allotrope of carbon \cite{zhang_penta-graphene:_2015}.

Previous force field studies have suggested that a rhombic phase is the most stable 
at ambient pressures \cite{zangi_monolayer_2003, koga_phase_2005}. 
Our own force field studies with either TIP4P/2005 \cite{vega_what_2008} or SPC/E \cite{berendsen_missing_1987} 
indeed
find that the rhombic structure has the lowest enthalpy (Fig. S8) \cite{si}. 
However with DFT it is considerably less stable than the hexagonal and CT structures. 
Similarly a recent DFT study concluded that a square structure is more stable
than any hexagonal structure \cite{corsetti_new_2015}.
Here we find that our most stable square structure has
a higher enthalpy by 43 meV/$\text{H}_\text{2}\text{O}$ than the hexagonal phase at ambient pressure.
In the SI we trace this difference to the different computational
setups \cite{si}.
As shown in the SI we are confident that
the hexagonal and CT monolayer ice structures are indeed more stable 
than any square ice structure at the low pressure limit.

\begin{figure}[h!]
\begin{center}
\includegraphics[width=8cm]{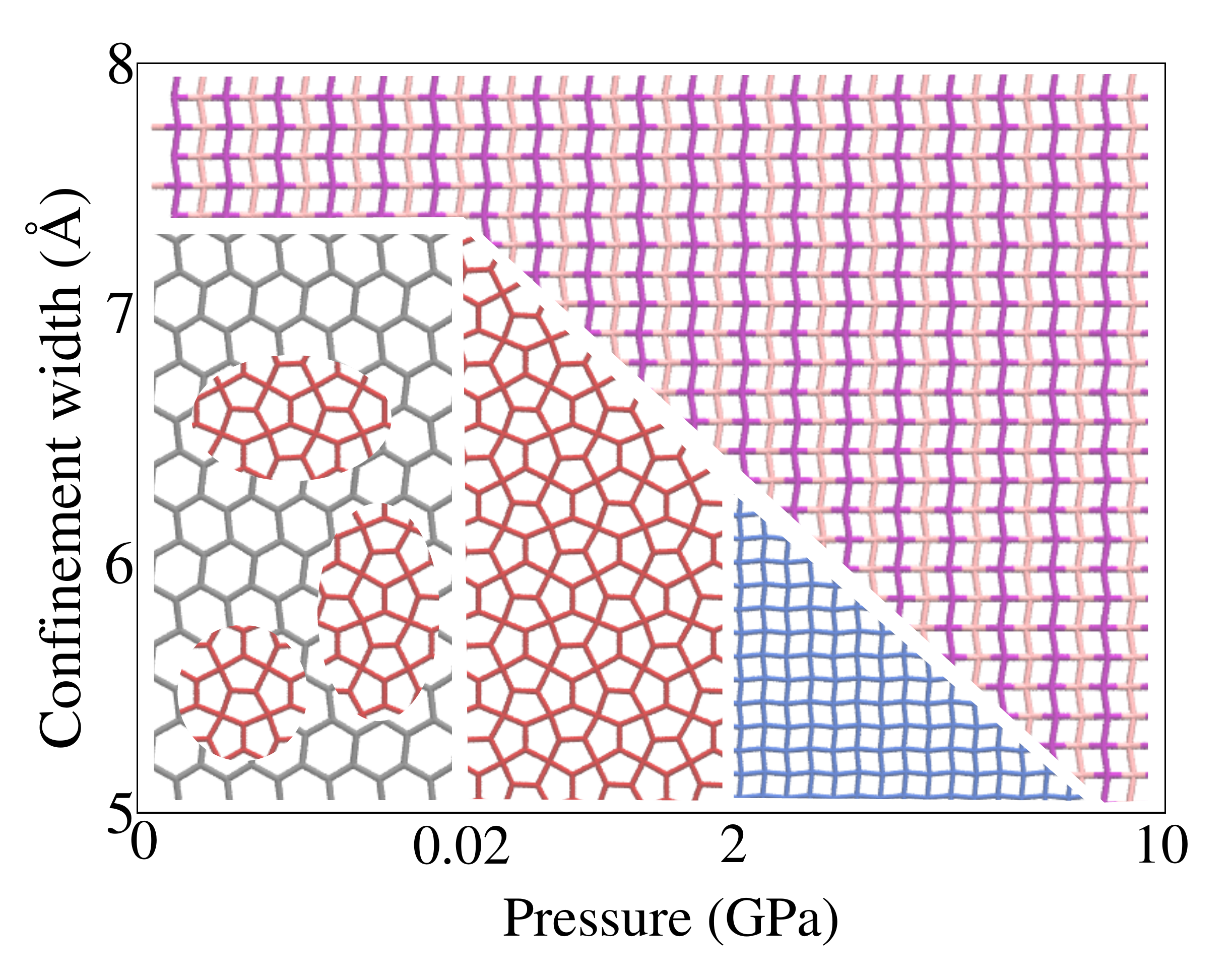}
\end{center}
\caption{Sketch of the phase diagram of monolayer ice with respect to lateral pressure and confinement width.
For clarity the pressure axis has an artificial scale with the various pressures
indicated corresponding to the transition pressures obtained at 6 \AA~confinement. 
Note that at ambient pressure the enthalpies of the hexagonal and Cairo tiled phases
are essentially degenerate but by \textit{ca.} 0.02 GPa the Cairo tiled phase is
more clearly favored.
See Fig. S7 for more results of enthalpies with respect to confinement width.
}
\label{figure3}
\end{figure}

At pressures beyond 2 GPa
higher density phases obeying the Bernal-Fowler-Pauling (BFP) 
ice rules are identified more frequently in
the structure searches. This includes a flat square phase (f-SQ) and
a buckled rhombic phase (b-RH).
Both phases consist of four-fold coordinated water molecules                       
with two donor HBs and two acceptor HBs.
The f-SQ structure has a $p4gm$ wall-paper group symmetry where
the dipoles of the water molecules are distributed on two orthogonal antiferroelectric 
sublattices (Fig. \ref{figure1}c).
The HB network of the b-RH phase is similar to f-SQ but 
it is buckled and has a higher lateral density (Fig. \ref{figure1}d).
The relative stability of the f-SQ and b-RH structures
in the 2-4 GPa regime depends sensitively on the confinement width (Fig. \ref{figure2}).
At 6.0 \AA~f-SQ is more stable while at 6.5~\AA~b-RH has the
lower enthalpy.
Beyond 4 GPa the b-RH phase is more stable than any other structure identified.
Several other metastable structures belonging to the b-RH family with different hydrogen bond ordering
have also been observed.
However, since a more delicate discussion of hydrogen ordering is beyond the scope of this study,
we only show one of the most stable members of the b-RH family at the
pressure and confinement conditions considered (Fig. \ref{figure1}d).

We also identified a second metastable square phase which
we dub ``b-SQ'' because of its buckled ``basketweave''-like pattern of HBs (Fig. \ref{figure1}e).
Its lattice structure resembles a 2D projection of bulk ice VIII but it is unique in
that the two sublayers are hydrogen bonding with each other.
The energy of the b-SQ phase is higher than the most stable phases identified,
however because of its unique hydrogen bonding arrangement and as 
it might be possible to observe it in systems where the substrate has
a square lattice we feel
it is worth reporting.

Beyond the phase behavior of monolayer ice confined within the 6.0 \AA~to 6.5 \AA~regime, 
we also explored a broader range of confinement widths at different pressures. 
These additional calculations support the validity of the conclusions reached but also show that
there is scope for altering the relative enthalpies of the various
phases by tuning the confinement.
The phase diagram for monolayer water with respect to lateral pressure
(0-10 GPa) and confinement width (5-8 \AA) that emerges from these calculations is shown 
schematically in Fig. \ref{figure3}.
At small confinements and low pressures ($<\sim$ 0.1 GPa) the hexagonal and the CT phases are preferred.
Increasing the pressure at small confinement widths results in a sequence of phase transitions from
the pentagonal to the square and the rhombic phases.
For larger confinement widths the b-RH phase is generally favored.

The f-SQ phase is found to be stable in the 2-4 GPa regime
at 6.0 \AA~confinement.
The structure of the f-SQ phase and the approximate pressure at which it appears
are
consistent with the recent experimental observation of a square ice phase in
graphene nanocapillaries \cite{algara-siller_square_2015}.
This lends some support to the predictions made here and suggests that it might also
be possible to observe the other structures predicted, by for example controlling
the density of water inside the nanocapillaries or with another 2D material
with lower interlayer adhesion.
In addition to TEM which has already been used,
techniques such as scanning tunneling microscopy
and atomic force microscopy might be able to further substantiate 2D ice lattices and hydrogen 
ordering in the future \cite{zhang_real-space_2013, guo_real-space_2014}.
In Fig. S6 we also show that different 2D ice monolayer structures have quite different 
vibrational properties \cite{si}.
For example the hexagonal and the CT phase have vibrational modes around 3700 $cm^{-1}$ 
due to dangling OH groups and lower 
frequency stretching modes $<$ 3000 $cm^{-1}$ arising from strong HBs.
The stretching regions also have quite different total widths for the different phases and
the bending mode of the b-SQ phase is softer than the other phases 
by \textit{ca.} 120 $cm^{-1}$.
Therefore it should also be possible to discriminate one phase from another with vibrational
spectroscopy.

\begin{figure}[h!]
\begin{center}
\includegraphics[width=8cm]{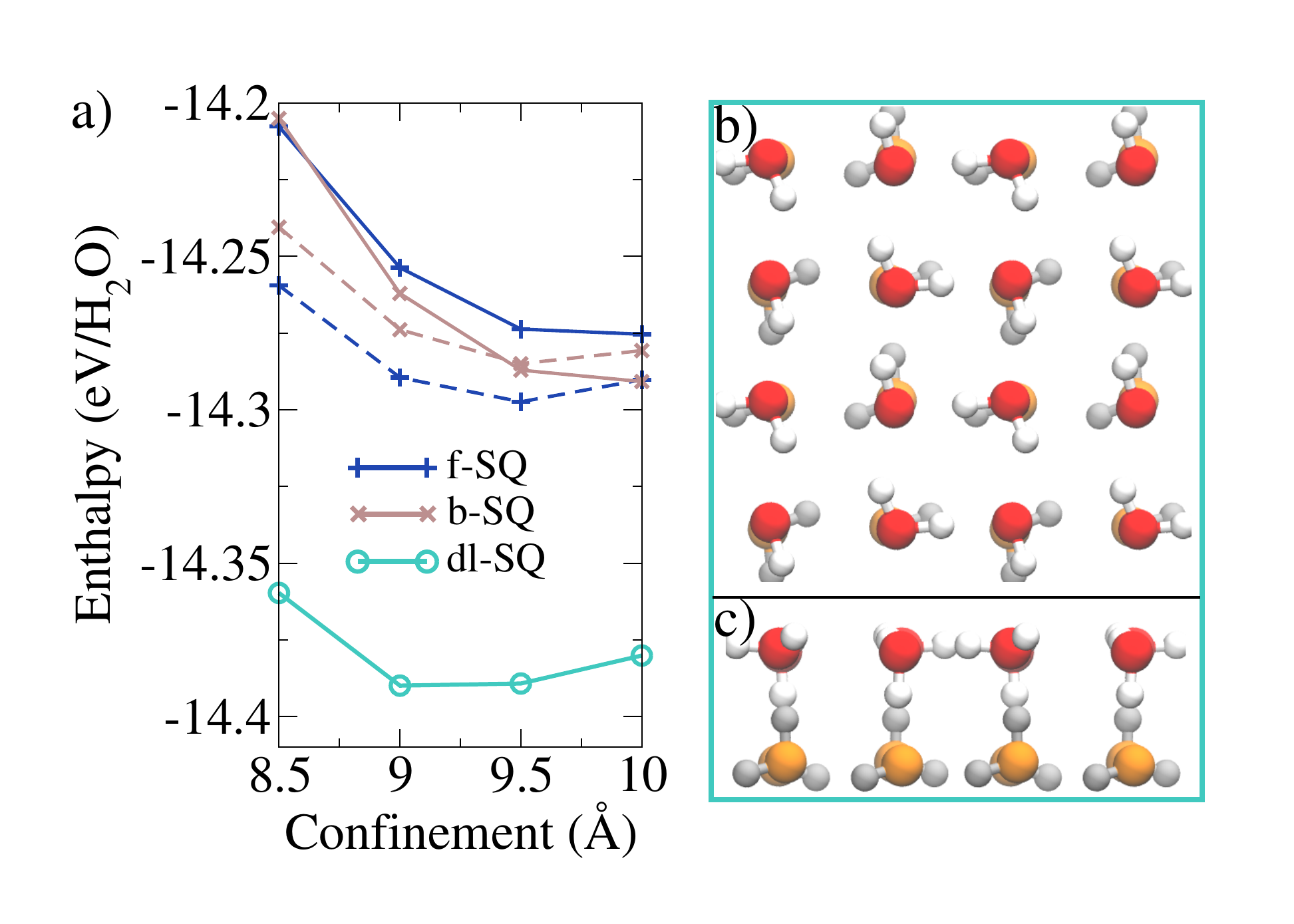}
\end{center}
\caption{Double layer ice energies and structures. 
(a) Enthalpy per water molecule for different double layer 
square ice structures as a function of confinement width at 2 GPa.
The solid lines for the f-SQ and b-SQ phases represent an AA stacking whereas the 
dashed lines are for AB stacking. 
(b,c) Structure of the dl-SQ phase.
Other AA and AB stacking double layer ice structures are shown in the SI \cite{si}.
}
\label{figure4}
\end{figure}

Multi-layer square ice was also observed experimentally with a structure in which oxygen atoms
are located directly on top of each other, 
suggesting an AA stacking square phase \cite{algara-siller_square_2015}.
However, the force field simulations performed as part of the experimental study obtained an 
AB stacking arrangement. 
The difference was attributed to a possible inaccuracy of the force field models used. 
In order to clarify this difference between experiment and theory, 
we calculated the enthalpy for different confinement widths at 2 GPa 
for a number of AA and AB stacking structures (Fig. \ref{figure4}a).
In agreement with the force field simulations we find AB stacking is preferred for the f-SQ phase.
However the other square structure identified in this study, b-SQ, is more stable with AA stacking, which agrees 
with experiment. 
The main interaction between the layers in these structures is van der Waals, with
the HB networks within the layers remaining unperturbed.
We find, however, that these stacked van der Waals bonded structures are considerably less 
stable than double layer structures with interlayer hydrogen bonds, 
the most stable of which we dub ``dl-SQ''.
The structure of dl-SQ is shown in Fig. \ref{figure4}b,c.
It is AA stacked and as can be seen from Fig. \ref{figure4}a it is \textit{ca.} 80 meV/$\text{H}_\text{2}\text{O}$
more stable than any of the stacked van der Waals bonded double layer structures.
Thus we conclude that the AA stacked double layer
square structure observed in experiments is unlikely to be a
van der Waals bonded layered structure but rather
one stabilised by interlayer hydrogen bonds.

In summary, monolayer ice phases and their phase transitions in confinement 
have been studied with DFT and a random structure search approach.
At ambient pressure we have predicted hexagonal and pentagonal Cairo tiled structure, which 
are similar in enthalpy and more stable than other structures.
The CT structure becomes more stable than the hexagonal structure when lateral pressure is applied.
Upon increasing the pressure to above about 2 GPa high density square and rhombic phases are observed.
Looking forward a complete description of the temperature dependent phase diagram of 2D water is desirable.
Experimentally it would be interesting to explore a broader range of temperatures, water densities, and confining
materials.
From the computational perspective, it would be desirable to explore 2D ice at higher temperature with electronic structure methods,
such as \textit{ab initio} molecular dynamics.
In addition, in light of the small ring sizes and relatively short intermolecular separations,
2D ice also provides further opportunities for investigating collective proton 
quantum dynamics \cite{li_quantum_2010, drechsel-grau_quantum_2014, meng_direct_2015}.

\section*{Acknowledgements}
J.C. and
A.M. are supported by the European Research Council
under the European Union's Seventh Framework Programme
(FP/2007-2013) / ERC Grant Agreement number
616121 (HeteroIce project). A.M is also supported by the Royal Society
through a Royal Society Wolfson Research Merit Award.
G.S. is supported by the TOUCAN EPSRC grant (EP/J010863/1).
C.G.S is supported by the Royal Society (UF100144).
We are also grateful for computational resources to the
London Centre for Nanotechnology, UCL Research Computing,
and to the UKCP consortium (EP/ F036884/1) for access
to Archer.


\bibliography{ref}

\end{document}